**Insensitivity of magnetic anomalies in $Sr_3NiPtO_6$ to positive and negative pressures**

S. Narayana Jammalamadaka, Niharika Mohapatra, Karthik K. Iyer and
E. V. Sampathkumaran[*]

*Tata Institute of Fundamental Research, Homi Bhabha Road, Colaba, Mumbai – 400005, India*

## ABSTRACT

The compound $Sr_3NiPtO_6$, belonging to a $K_4CdCl_6$-derived rhombohedral structure, has been reported not to exhibit magnetic ordering at least down to 1.8 K, despite a relatively large value of paramagnetic Curie temperature. This is attributable to geometrical frustration. Here we report the results of our efforts to gradually replace Sr by Ba and to probe the influence of positive (external) and negative (chemical) pressure on the magnetic behavior of this compound. In the Ba substituted series, single phase is formed up to x= 1.0 with Ba substituting for Sr. The magnetic properties of the parent compound in the entire temperature range of investigation are not influenced at all in any of the compositions studied as well as under external pressure (investigated up to 10 kbar). Spin-liquid-like heat-capacity behavior (finite linear term) is observed even in Ba-substituted specimens. Thus, the magnetic anomalies of this compound are quite robust.

[*]Corresponding author: Tel: +91 22 22782437; fax: +91 22 22804610.
E-mail address: sampath@mailhost.tifr.res.in (E.V. Sampathkumaran)

## 1. Introduction

The investigation of topologically frustrated magnetism is one of the important current topics in the field of magnetism. In particular, the spin-liquid phenomenon proposed for some geometrically frustrated systems [see, for example, Refs. 1-3] is a fascinating one. Recently, the compounds of the type $(Sr,Ca)_3XYO_6$ (X, Y= a metallic ion, magnetic or non-magnetic), crystallizing in the $K_4CdCl_6$-derived rhombohedral structure (space group: $R\bar{3}c$) have been attracting a lot of attention [See, references 4-10, and articles cited therein] with respect to various manifestations of geometrical frustration, as the crystal structure is made up of chains of X and Y ions (running along $c$-direction) arranged hexagonally in the basal plane forming a triangular lattice. These chains are separated by Sr (or Ca) ions. It is found that, if X and/or Y carry magnetic moment, interchain interaction is antiferromagnetic and therefore triangular arrangement of magnetic ions in the basal plane results in magnetic frustration. It was found [10] that the compound $Sr_3NiPtO_6$ (Refs. 11, 12) does not order magnetically (at least down to 1.8 K), attributable to geometrical frustration, despite the fact that the magnitude of the paramagnetic Curie temperature ($\theta_p$) is large (about -25 K). This is interesting considering that spin-polarized band structure calculations [13] reveal that the Ni carries a spin magnetic moment of about 1.7 $\mu_B$ (with Pt carrying negligibly small moment). In this compound, Ni is in 2+ state with high-spin $d^8$ configuration (S= 1 state) and Pt is in 4+ state with low-spin $d^6$ configuration. Therefore, if the absence of magnetic ordering is due to spin-liquid phenomenon, it would be exciting, as such a phenomenon is not so-commonly reported among S= 1 systems. Even in the case of $NiGa_2S_4$ [Ref. 1], there is a recent claim [14] that Ni moments freeze out below 10 K. Thus, the observed low-temperature property of $Sr_3NiPtO_6$ is of great importance and therefore it is of interest to get more information with respect to the magnetic properties of this compound. We carried out the present investigation primarily to see whether the magnetic property of this compound can be altered by negative and positive pressures by chemical doping and external pressure. For this purpose, we have attempted substitution of Sr by Ba, in addition to carrying out high pressure experiments. It is found that Ba could partially replace Sr and that the magnetic property is essentially retained for all the compositions studied.

## 2. Experimental details

Polycrystalline samples of $Sr_{3-x}Ba_xNiPtO_6$ (0, 0.5, 1.0 and 1.5) were synthesized *via* a solid state route. $SrCO_3$ (99.994%), NiO (99.995%), $Pt_2O_2.H_2O$ (99.999%) and $BaCO_3$ (99.999%) were thoroughly mixed in an agate mortar and calcined in air at 800°C. Then the preheated powder was reground, pressed, and subsequently sintered at 1000°C for 9 days with three intermediate grindings. The samples were characterized by powder x-ray diffraction (XRD) with Cu $K_\alpha$ radiation. Further characterization of the samples was carried out by scanning electron microscope (SEM) by energy dispersive X-ray analysis (EDXA) and back-scattered electron (BSE) imaging. Dc magnetic susceptibility ($\chi$, defined as $M/H$ where $M$ is the magnetization and $H$ is the applied magnetic field) studies (1.8 – 300 K) were performed by a commercial (Quantum Design) superconducting quantum interference device. The dc magnetization under pressure for the parent compound was measured (1.8 – 300 K) in a hydrostatic pressure medium (daphne oil) up to 10 kbar using a commercial (EasyLab Technologies Ltd., UK) pressure cell with the same magnetometer. Isothermal magnetization behavior at 5K was tracked with a commercial (Oxford Instruments) vibrating sample magnetometer up to 100 kOe. In some cases, we have also measured heat capacity (C) (1.8-30 K) with a commercial (Quantum Design) physical properties measurements system in zero magnetic field and in 50 kOe.

**3. Results and discussions**
*3.1. Solid solution behavior*

All the XRD diffraction lines (see figure 1) of $Sr_{3-x}Ba_xNiPtO_6$ for $x$ = 0.5 and 1.0 could be indexed to $K_4CdCl_6$-type rhombohedral structure. In addition to this phase, some weak lines attributable to a small amount of BaO phase could be detected for $x$= 1.5 composition only. As expected for Ba substitution for Sr, there is a gradual expansion of the unit-cell with increasing Ba concentration (see table 1 for lattice constants). This is convincingly demonstrated by showing the data in the higher angle side in an expanded form in figure 1. We have performed Reitveld analysis for single-phase compositions to ensure proper stoichiometry and site occupation and the refinement parameters ($R_{wp}$, $R_p$ and $\chi^2$ with usual meanings in such an anlaysis), included in figure 1, fall in a satisfactory range. We have also carefully analyzed SEM/BSE images and we found that the material is homogeneous across entire specimen. The stoichiometry in all these compositions has been confirmed with the EDXA. But the BaO phase is not clearly resolvable for $x$= 1.5, possibly due to homogeneously mixed particles of very small size.

At this juncture, we would like to mention about the solid solution behavior of the series $Sr_{3-x}Ca_xNiPtO_6$ ($x$ = 0, 0.5, 1.0, and 1.5), which we attempted to synthesize with the aim of inducing positive chemical pressure. XRD patterns, Rietveld analysis of the XRD patterns, BSE images, and EDAX analysis revealed that NiO phase gradually precipitates with increasing $x$. About 10 atomic percent of Ca occupies Ni as known earlier for this family [15]. Consistent with this, we found that the lattice expands with $x$ in this series. The ionic radius of $Ca^{2+}$ for the same coordination being more by about 0.3 Å compared to that of divalent Ni, preference of Ca for Ni site causes lattice expansion. Thus, for instance, the main phase for the nominal composition, $x$= 0.5 in $Sr_{3-x}Ca_xNiPtO_6$, is found to correspond to $(Sr_{2.5}Ca_{0.4})(Ni_{0.9}Ca_{0.1})PtO_6$. As these specimens are not single-phase, we do not present the data in detail in this article. We have confirmed partial occupation of Ca at the Ni site by preparing the single phase compound with a composition, $Sr_3Ni_{0.9}Ca_{0.1}PtO_6$.

*3.2. Magnetization behavior*

The results of magnetization measurements of $Sr_{3-x}Ba_xNiPtO_6$ are shown in figure 2. As reported earlier [8], the $\chi$, measured in a field of 5 kOe, for $Sr_3NiPtO_6$ exhibits Curie-Weiss behavior at high temperatures, say, above 150 K, below which there is a gradual deviation from linearity with decreasing temperature. At low temperatures, say, below about 25 K, there is a sluggish temperature dependence as though there is a tendency to flatten. If $\chi$ is measured in the presence of high fields, say at 50 kOe, the temperature dependence gets further weakened. In some specimens, a weak maximum around 25 K is observed [10] as shown in figure 3. This means that the weak low-temperature upturn observed for some specimens is not intrinsic, and the compound in fact does not order magnetically down to 1.8 K. We did not find any evidence for the bifurcation of the $\chi(T)$ curves for the zero-field-cooled and field-cooled conditions of the specimen, even when measured in low fields (100 Oe). The effective moment ($\mu_{eff}$) obtained from the Curie-Weiss region is about 3.3 $\mu_B$ per formula unit and the value of $\theta_p$ is about -11 K in the specimen employed here. The magnitude of $\theta_p$ reported here is lower compared to that reported in Ref. 10, implying sample dependence of this parameter possibly due to orientation effects in such low-dimensional systems. Qualitatively speaking, the magnetic behavior is very weakly affected with the gradual replacement of Sr by Ba. Curie-Weiss behavior of $\chi$ over a wider temperature range as in parent compound is retained and there is a negligible change in $\mu_{eff}$ (within experimental error). The magnitude of $\theta_p$ also changes only marginally, as though the exchange interaction is not influenced in any significant manner. There is no evidence for long range ferro/antiferro magnetic ordering down to 1.8 K. Thus, negative chemical pressure does not bring out any notable change in the magnetic properties, despite a volume expansion of about 3 to 4%.

As remarked earlier, it turned out that, in the Ca substituted specimens, there is a lattice expansion apart from the fact that these specimens contain some amount NiO phase. Hence we could not investigate lattice compression effect on magnetism through this solid solution. Therefore, we carried out high pressure magnetization studies till 10 kbar. As shown in figure 3, the $\chi$(T) behavior as well as M(H) behavior are the same as those at ambient pressure. This implies that positive pressure (till the pressure range employed) does not alter the magnetic properties noticeably.

Incidentally, we have also measured magnetization of the Ca specimens and it was found that the low-temperature behavior (absence of magnetic ordering) is found to be the same as the parent compound, despite the fact that the presence of NiO phase. NiO has been known to be a well-celebrated antiferromagnet. However, the $\chi$ values were found to get decreased with increasing NiO content in this matrix. We therefore raise a question whether in this composite form (that is, micron-sized NiO dispersed in the matrix of this Sr/Ca oxide) the magnetism of NiO is quenched. This behavior of NiO is fascinating, as it is well known that even in the nanoform (about 20 nm) NiO retains large moments and coercivities [16].

*3.3. Heat-capacity behavior*

We have measured heat capacity on selected compositions (see figure 4) in zero field and in the presence of 50 kOe. It is not possible to subtract phonon contribution as no satisfactory reference for lattice part exists. It is however rather straightforward to draw the inferences even from the raw data. The results on the parent compound, reproduced in figure 4, have been reported earlier by us [10]. As in the parent compound, there is no evidence for any feature attributable to magnetic ordering down to 1.8 K in the data for Ba compositions. Even an application of a magnetic field as large as 50 kOe does not alter the behavior, including the magnitudes of C in any significant way, even for these chemically doped compositions. This insensitivity to application of magnetic fields is noteworthy, ruling out antiferromagnetic or spin-glass ordering. However, the weak upturn present in C/T at low temperatures for $x$= 0.0 is suppressed for $x$= 1.0. C/T varies quadratically with T below about 10 K, with a non-vanishing linear term (20 mJ/mol K$^2$ for Ba specimen in figure 4) despite the fact that these materials are insulators. This 'gapless' behavior is similar to that observed in systems proposed for spin-liquid behavior in the literature [1, 2].

**4. Conclusion**

We have probed the solid solution formation in the Ba substituted Sr$_3$NiPtO$_6$ and the magnetic behavior of this compound due to external pressure and negative chemical pressure in this compound. Negative chemical pressure and external pressure on Sr$_3$NiPtO$_6$ do not bring out any significant change in the magnetic properties. It is thus fascinating to note that no magnetic ordering is observed in any of these substituted oxides, despite the fact that the magnitude of $\theta_p$ is reasonably large. Spin-liquid-like heat-capacity behavior (finite linear term) is observed even in Ba-substituted specimens. Thus, the non-magnetic ground state of $Sr_3NiPtO_6$ is quite robust.

**Acknowledgements**

We would like to thank B. A. Chalke for her help during SEM/BSE measurements.

**References**
[1]  For a Ni-based system, see, S. Nakasutji, Y. Nambu, H. Tonomura, O. Sakai, S. Jonas, C. Broholm, H. Tsunetsugu, Y. Qiu, and Y. Maeno, Science 309 (2005) 1697-1700
[2]  For a Ir-based system, see, Y. Okamoto, M. Nohara, H. Aruga-Katori, and H. Takagi, Phys. Rev. Lett. 99 (2007) 137207.
[3]  For Tb-systems, see, I. Mirebeau, I.N. Goncharenko, P. Cadavez-Peres, S.T. Bramwell, M.J.P. Gingras, and J.S. Gardner, Nature 420 (2002) 54-57; J.S. Gardner, B.D. Gaulin, A.J. Berlinsky, and P. Waldron, Phys. Rev. 64 (2001) 224416.
[4]  See the reviews K.E. Stitzer, J. Darriet, and H.-C. zur Loye, Curr. Opin. Solid State Mater. Sci. 5 (2001) 535; G.V. Bazuev, Russ. Chem. Rev. 75 (2006) 749-763
[5]  J. Darriet, F. Grasset, and P. D. Battle, Mater. Res. Bull. 32 (1997) 139; T. N. Nguyen and H. C. zur Loye, J. Solid State Chem. 117 (1995) 300-308
[6]  H. Kageyama, K. Yoshimura, K. Kosuge, H. Mitamura, and T. Goto, J. Phys. Soc. Jpn. 66 (1997) 1607-1610.
[7]  S. Niitaka, K. Yoshimura, K. Kosuge, M. Nishi, and K. Kakurai, Phys. Rev. Lett. 87 (2001) 177202.
[8]  A. Maignan, A. C. Masset, C. Martin, B. Raveau Eur. Phys. J. B 15 (2000) 657-663; S. Rayaprol, Kausik Sengupta, and E.V. Sampathkumaran, Solid State Commun. 128 (2003) 79-84 and references therein.
[9]  Niharika Mohapatra, Kartik K Iyer, Sudhindra Rayaprol, S. Narayana Jammalamadaka, and E.V. Sampathkumaran, J. Phys.: Condens. Matter 20 (2008) 255247.
[10] Niharika Mohapatra, Kartik K Iyer, Sudhindra Rayaprol, and E.V. Sampathkumaran, Phys. Rev. B 75 (2007) 214422 and references therein.
[11] T.N. Nguyen, D.M. Giaquinta, and H.-C. zur Loye, Chem. Mater. 6 (1994)1642; J. B. Claridge, R. C. Layland, W. H. Henley, and H. C. zur Loye, Chem. Mater. 11 (1999)1376-1380.
[12] G.V.Vajenine, R. Hoffmann, and H.-C. zur Loye, Chem. Phys. 204 (1996) 469-478.
[13] S.K. Pandey and K. Maiti (private communication).
[14] H. Takeya, K. Ishida, K. Kitagawa, Y. Ihara, K. Onuma, Y. Maeno, Y. Nambu, S. Nakasutji, D.E. MacLaughlin, A. Koda, and R. Kadono, Phys. Rev. B. 77 (2008) 054429.
[15] R.C. Layland and H.-C. Loye, J. Alloys and Compd. 299 (2000) 118-125; K.E. Stitzer, W.H. Henley, J.B. Claridge, H.-C. Loye, and R.C. Layland, J. Solid State Chem. 164 (2002) 220-229; E.V. Sampathkumaran, Z. Hiroi, S. Rayaprol, and Y. Uwatoko, J. Magn. Magn. Mater. 284 (2004) L7-11.
[16] See, for example, R.H. Kodama, Salah A. Makhlouf, and A.E. Berkowitz, Phys. Rev. Lett. 79 (1997) 1393-1396.

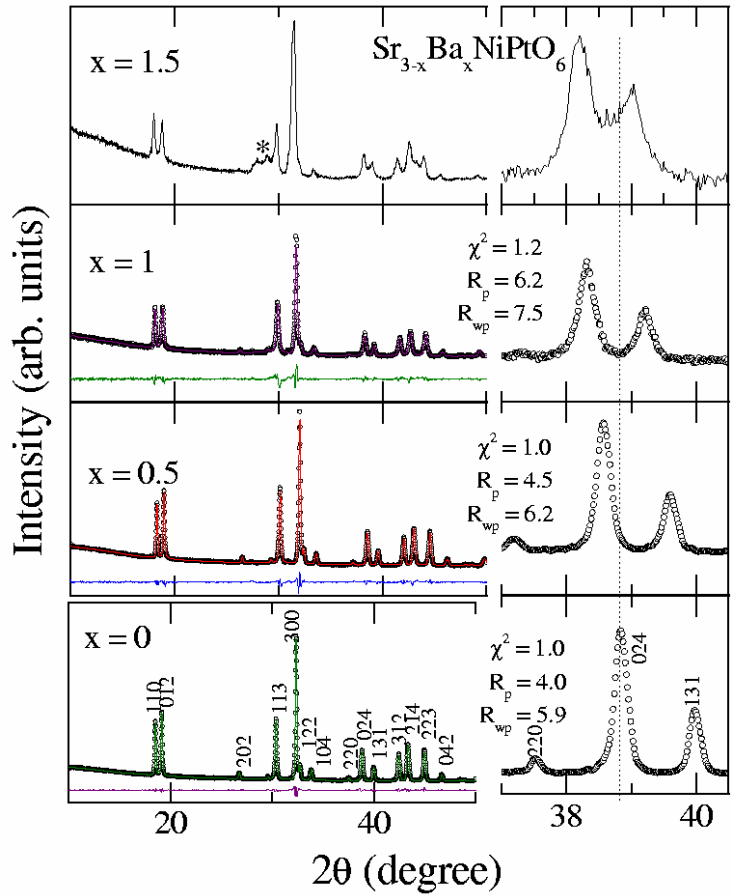

Fig. 1. (Color online) X-ray diffraction patterns of $Sr_{3-x}Ba_xNiPtO_6$ compounds. The asterisk indicates the BaO phase. In the case of $x=$ 0.0, 0.5 and 1.0), the symbols are experimental data points, while the lines through them are obtained by Reitveld fitting; the difference spectra between the experimental and fit spectra are also included at the bottom in respective panels. Some diffraction lines are shown in an expanded form to show that the lattice expands with increasing Ba concentration.

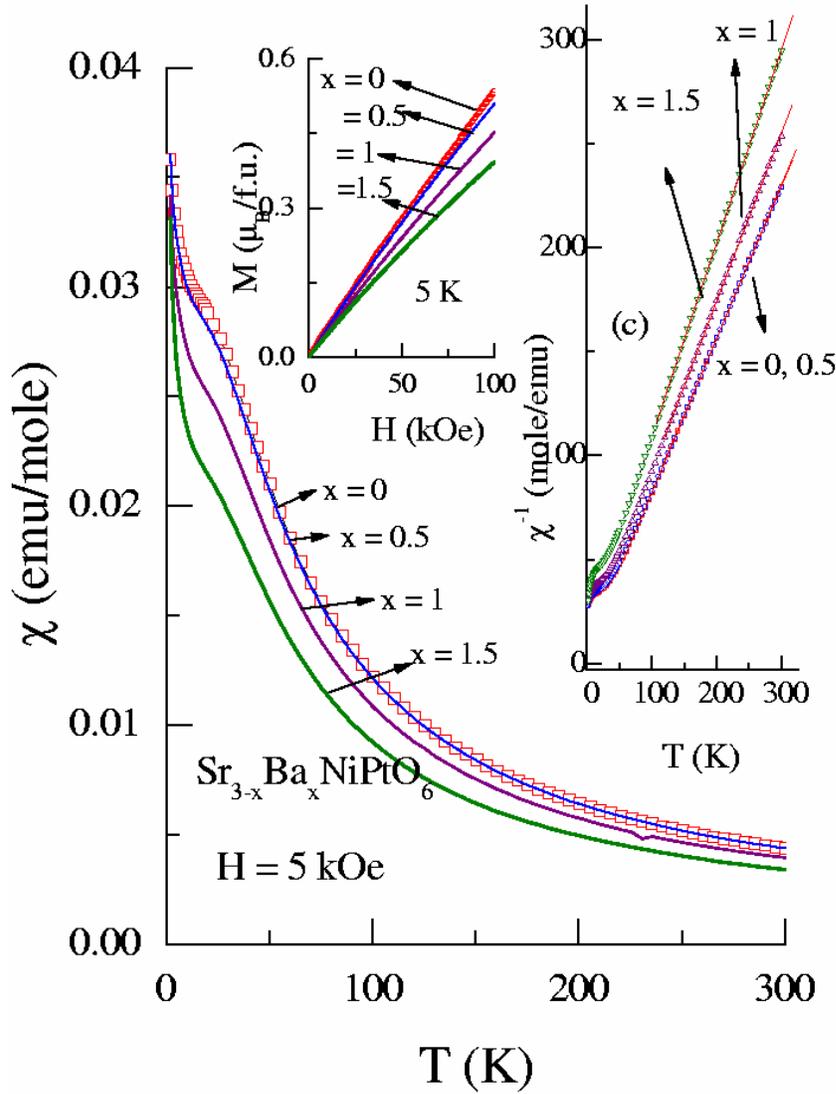

Fig. 2. (Color online) Dc magnetic susceptibility ($\chi$) as a function of temperature for $Sr_{3-x}Ba_xNiPtO_6$ (x = 0, 0.5, 1.0 and 1.5) obtained in a field of 5 kOe. Isothermal magnetization (M) upto 100 kOe at 5 K and inverse $\chi(T)$ versus T are shown in the insets. The continuous lines in inverse $\chi(T)$ plots represent the high temperature Curie-Weiss region.

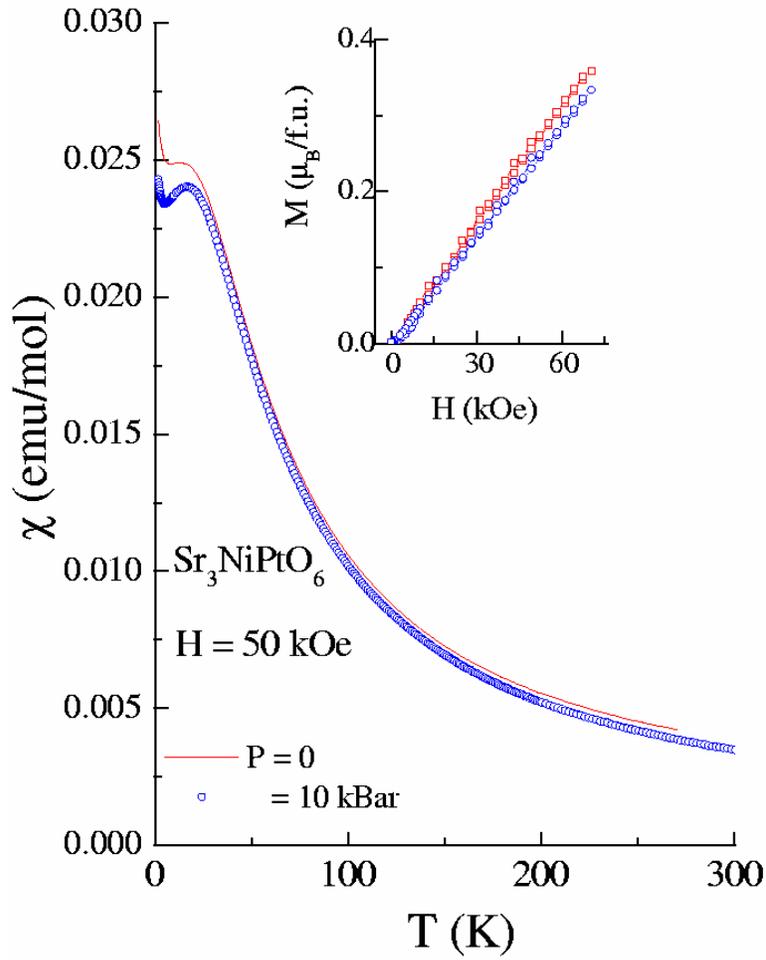

Fig. 3. (Color online) Dc magnetic susceptibility ($\chi$) as a function of temperature for $Sr_3NiPtO_6$ obtained in a field of 50 kOe at ambient pressure and 10 kbar measured at ambient pressure. The inset shows isothermal magnetization data at 1.8 K.

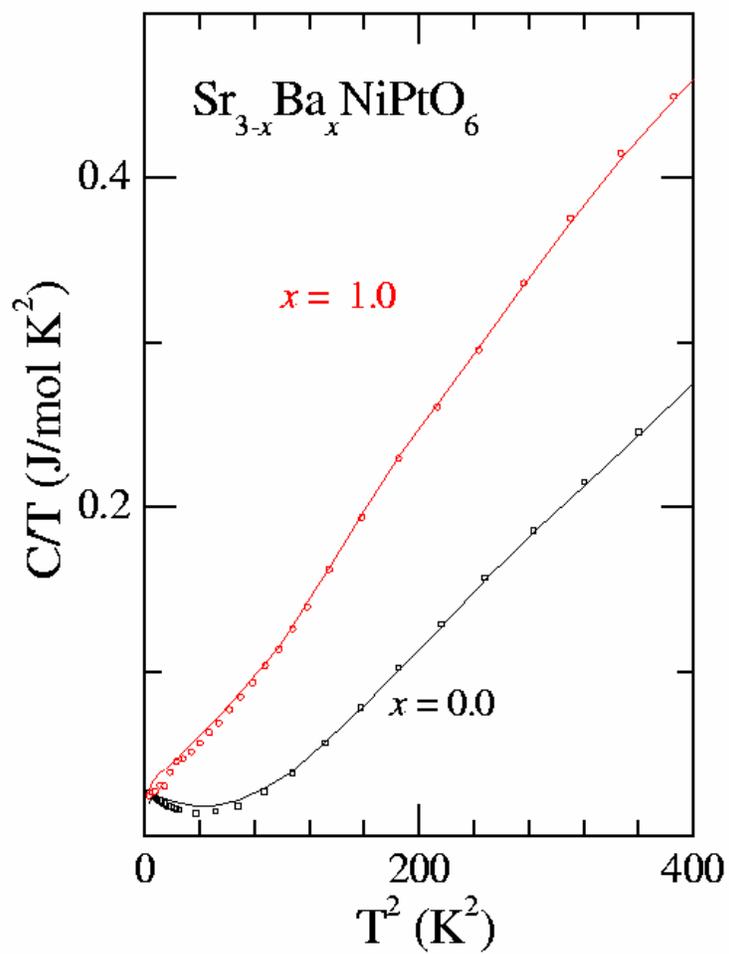

Figure 4:
(Color online) Heat capacity divided by temperature versus square of temperature for $Sr_3NiPtO_6$ and $Sr_2BaNiPtO_6$. While the data points correspond to zero field, the lines shown are obtained in a field of 50 kOe after omitting data points.

**Table 1**
Lattice parameters (*a* and *c*), unit-cell volume (V), paramagnetic Curie temperature ($\theta_p$), effective moment ($\mu_{eff}$) for $Sr_{3-x}Ba_xNiPtO_6$ ($x$ = 0, 0.5, 1 and 1.5).

| X | $a$ ($\pm0.002$Å) | $c$ ($\pm0.002$Å) | V ($\pm0.5$Å$^3$) | $\theta_p$ ($\pm 2$ K) | $\mu_{eff}$ ($\pm 0.1\mu_B$) |
|---|---|---|---|---|---|
| 0.0 | 9.572 | 11.172 | 886 | -11 | 3.28 |
| 0.5 | 9.699 | 11.270 | 918 | -13 | 3.30 |
| 1.0 | 9.781 | 11.310 | 937 | -15 | 3.14 |
| 1.5 | 9.822 | 11.334 | 946 | -18 | 2.94 |